\documentclass[twocolumn]{aastex63}
\usepackage{verbatim}
\usepackage{amsmath}
\let\tablenum\relax
\usepackage[detect-all]{siunitx}
\DeclareUnicodeCharacter{2212}{-}
\usepackage{xspace}
\usepackage{threeparttable}
\usepackage{graphicx}
\newsavebox{\measurebox}
\usepackage[caption=false]{subfig}
\usepackage{lineno}
\usepackage{booktabs}
\usepackage{multirow}
\usepackage{dashrule}

\usepackage{xcolor,ulem}
\newcommand\hlD{\bgroup\markoverwith{\textcolor{yellow}{\rule[-.5ex]{2pt}{2.5ex}}}\ULon}
 
\defcitealias{Harris1996}{H96} 	  	 
\newcommand{\harris}{\citetalias{Harris1996}\xspace}
\newcommand{\harrisp}{\citepalias{Harris1996}\xspace}
\newcommand{\astrosat}{\textit{AstroSat}}
\newcommand{\hst}{{\textit {HST}}\xspace}
\newcommand{\gaia}{{\textit{Gaia EDR3}}\xspace}
\newcommand{\omcen}{{$\omega$~Cen}\xspace}
\newcommand{\teff}{{$T_{\rm eff}$}\xspace}

\newcommand{\rad}{{$R$}\xspace}
\newcommand{\logg}{\ensuremath{\log g}\xspace}
\newcommand{\feh}{{\rm{[Fe/H]}}\xspace}
\newcommand{\afe}{{\rm{[$\alpha$/Fe]}}\xspace}

\newlength\replength
\newcommand\repfrac{.33}

\newcommand\rulewidth{.6pt}
\newcommand\tdashfill[1][\repfrac]{\cleaders\hbox to \replength{%
  \smash{\rule[\arraystretch\ht\strutbox]{\repfrac\replength}{\rulewidth}}}\hfill}

\newcommand\tdotfill[1][\repfrac]{\cleaders\hbox to \replength{%
  \smash{\raisebox{\arraystretch\dimexpr\ht\strutbox-.1ex\relax}{.}}}\hfill}

\received{}
\revised{}
\accepted{}

\submitjournal{ApJ Letters}

\shortauthors{Prabhu et al.}

\graphicspath{{./}{}}

\begin{document}

\title{Globular Cluster UVIT legacy Survey (GlobUleS) III. Omega Centauri in Far-Ultraviolet}

\correspondingauthor{Deepthi S. Prabhu}
\email{deepthi.prabhu@iiap.res.in}

\author[0000-0002-8217-5626]{Deepthi S. Prabhu}
\affiliation{Indian Institute of Astrophysics, Koramangala II Block, Bangalore-560034, India}
\affiliation{Pondicherry University, R.V. Nagar, Kalapet, 605014, Puducherry, India}

\author[0000-0003-4612-620X]{Annapurni Subramaniam}
\affiliation{Indian Institute of Astrophysics, Koramangala II Block, Bangalore-560034, India}

\author[0000-0002-0801-8745]{Snehalata Sahu}
\affiliation{Department of Physics, University of Warwick, Coventry CV4 7AL, UK}

\author[0000-0001-6812-4542]{Chul Chung}
\affiliation{Department of Astronomy \& Center for Galaxy Evolution Research, Yonsei University, Seoul 03722, Republic of Korea}

\author{Nathan W. C. Leigh}
\affiliation{Departamento de Astronom\'ia, Facultad Ciencias F\'isicas y Matematicas, Universidad de Concepci\'on, Av. Esteban Iturra s/n Barrio Universitario, Casilla 160-C, Concepci\'on, Chile}
\affiliation{Department of Astrophysics, American Museum of Natural History, Central Park West and 79th Street, New York, NY 10024-5192, USA 2}

\author[0000-0003-4237-4601]{Emanuele Dalessandro}
\affiliation{INAF-Astrophysics and Space Science Observatory Bologna, Via Gobetti 93/3 I-40129 Bologna, Italy}

\author[0000-0002-3680-2684]{Sourav Chatterjee}
\affiliation{Tata Institute of Fundamental Research, Mumbai 400005, India}

\author[0000-0002-8414-8541]{N. Kameswara Rao}
\affiliation{Indian Institute of Astrophysics, Koramangala II Block, Bangalore-560034, India}

\author[0000-0003-0155-2539]{Michael Shara}
\affiliation{Department of Astrophysics, American Museum of Natural History, Central Park West and 79th Street, New York, NY 10024-5192, USA 2}

\author[0000-0003-1184-8114]{Patrick C\^ot\'e}
\affiliation{Herzberg Astronomy and Astrophysics Research Centre, National Research Council of Canada, 5071 W. Saanich Road, Victoria, BC V9E 2E7, Canada}

\author[0000-0001-8182-9790]{Samyaday Choudhury}
\affiliation{Space Telescope Science Institute, 3700 San Martin Dr., Baltimore, MD 21218, USA}

\author[0000-0001-5812-1516]{Gajendra Pandey}
\affiliation{Indian Institute of Astrophysics, Koramangala II Block, Bangalore-560034, India}

\author{Aldo A. R. Valcarce}
\affiliation{Departamento de Física, Universidade Estadual de Feira de Santana, Av. Transnordestina, s/n, 44036-900, Feira de Santana, BA, Brazil}
\affiliation{Millennium Institute of Astrophysics MAS, Nuncio Monsenor Sotero Sanz 100, Of. 104, Providencia, Santiago, Chile}

\author[0000-0003-2952-3617]{Gaurav Singh}
\affiliation{Indian Institute of Astrophysics, Koramangala II Block, Bangalore-560034, India}

\author{Joesph E. Postma}
\affiliation{Department of Physics and Astronomy, University of Calgary, Calgary, AB T2N 1N4, Canada}

\author[0000-0003-4233-3180]{Sharmila Rani}
\affiliation{Indian Institute of Astrophysics, Koramangala II Block, Bangalore-560034, India}

\author[0000-0002-8304-5444]{Avrajit Bandyopadhyay}
\affiliation{Department of Astronomy, University of Florida, Gainesville, FL 32601, USA}

\author{Aaron M. Geller}
\affiliation{Center for Interdisciplinary Exploration and Research in Astrophysics (CIERA) and Department of Physics and Astronomy, Northwestern University, \\1800 Sherman Ave., Evanston, IL 60201, USA}

\author{John Hutchings}
\affiliation{Herzberg Astronomy and Astrophysics Research Centre, National Research Council of Canada, 5071 W. Saanich Road, Victoria, BC V9E 2E7, Canada}

\author[0000-0003-0350-7061]{Thomas Puzia}
\affiliation{Institute of Astrophysics, Pontificia Universidad Cat\'olica de Chile, Av. Vicuña MacKenna 4860, 7820436, Santiago, Chile}

\author[0000-0002-5652-6525]{Mirko Simunovic}
\affiliation{Subaru Telescope,National Astronomical Observatory of Japan, 650 N Aohoku Pl, Hilo, HI 96720, USA}

\author{Young-Jong Sohn}
\affiliation{Department of Astronomy \& Center for Galaxy Evolution Research, Yonsei University, Seoul 03722, Republic of Korea}

\author{Sivarani Thirupathi}
\affiliation{Indian Institute of Astrophysics, Koramangala II Block, Bangalore-560034, India}

\author{Ramakant Singh Yadav}
\affiliation{Aryabhatta Research Institute of Observational Sciences, Nainital, India}

\begin{abstract}
We present the first comprehensive study of the most massive globular cluster Omega Centauri in the far-ultraviolet (FUV) extending from the center to $\sim$~28\% of the tidal radius using the Ultraviolet Imaging Telescope aboard \astrosat. A comparison of the FUV-optical color-magnitude diagrams with available canonical models reveals that the horizontal branch (HB) stars bluer than the knee (hHBs) and the white dwarfs (WDs) are fainter in the FUV by $\sim$ 0.5 mag than model predictions.
They are also fainter than their counterparts in M13, another massive cluster. We simulated HB with at least five subpopulations including three He-rich populations with a substantial He enrichment of Y up to 0.43~dex, to reproduce the observed FUV distribution. We find the He-rich younger subpopulations to be radially more segregated than the He-normal older ones, suggesting an in-situ enrichment from older generations. The \omcen hHBs span the same \teff range as their M13 counterparts, but some have smaller radii and lower luminosities. This may suggest that a fraction of \omcen hHBs are less massive than those of M13, similar to the result derived from earlier spectroscopic studies of outer extreme HB stars.
The WDs in \omcen and M13 have similar luminosity-radius-\teff parameters and 0.44 - 0.46~M$_\odot$ He-core WD model tracks evolving from progenitors with Y = 0.4~dex are found to fit the majority of these. This study provides constraints on the formation models of \omcen based on the estimated range in age, [Fe/H] and Y (in particular), for the HB stars.
\end{abstract}


\section{Introduction} 

Galactic globular clusters (GCs) harbor stars hot enough to be significant emitters of ultraviolet (UV) light \citep[see][for detailed reviews]{Moehler2001,Moehler2010}. Studying these stars can help elucidate several problems in topics such as the late stages of low-mass stars' evolution \citep{Moehler2019}, stellar dynamics \citetext{eg., \citealt{Ferraro2012, leigh13}}, the “UV-upturn” seen in the spectra of elliptical galaxies \citep{GreggioRenzini1990,Dorman1993,Dorman1995} and so on. The identification of UV-bright stars is best done using UV images, as crowding due to populous cooler stars are suppressed in the central cores, at these wavelengths.

Omega Centauri (\omcen; or NGC\,5139), being the most massive GC in the Galaxy  \citetext{mass = 3.5 $\times$ $10^{6}$$M_{\odot}$; \citealt{Baumgardt2018}} contains the largest known population of very hot horizontal branch stars \citetext{HBs; \citealt{Dcruz2000}} and exotic blue straggler stars \citetext{BSSs; \citealt{Ferraro2006, Mucciarelli2014}}. Stars with a wide range in metallicity ($-$2.2 $\lesssim$ \feh $\lesssim$ $-$ 0.6 dex) and helium (He) abundance (Y up to 0.4~dex) have been reported through 
spectroscopic measurements \citep[and references therein]{Moehler2011, Monibidin2012, Latour2021} and through isochrone-fitting and population synthesis of color-magnitude diagrams \citetext{CMDs; \citealt{Norris2004, Piotto2005, Lee2005, JooLee2013, Tailo2016}}. The presence of He-core white dwarfs (WDs) has also been suggested in the studies of \citet{Calamida2008} and \citet{Bellini2013}.

The previous far-UV (FUV) study of this cluster was conducted decades back using the Ultraviolet Imaging Telescope \citetext{UIT, \citealt{Landsman1992, Whitney1994,Whitney1998}} and the \textit{Hubble Space Telescope} \citetext{HST, \citealt{Dcruz2000}}. However, these datasets are incomplete due to the limited spatial resolution of UIT ($\sim$~3$''$) and field of view (FOV) of HST/WFPC2.

In this Letter, we present the first comprehensive FUV investigation of \omcen extending from its center to $\sim$~28\% of the tidal radius, r$_{t}$ = 48$'$ \citep[2010 edition, hereafter \harris]{Harris1996}, carried out using the Ultra Violet Imaging Telescope (UVIT) onboard \astrosat. For the first time, we detect populations of HBs and WDs which are anomalously fainter in the FUV band as compared to theoretical models as well as their counterparts in another massive GC, M13.


\section{Observations and Data Reduction} \label{obs}

\omcen was observed as a part of the Globular Cluster UVIT Legacy Survey \citetext{GlobULeS, \citealt{Sahu2022}} on 24 January 2021 in two FUV filters, F148W and F169M, covering the entire 28$\arcmin$ diameter FOV of the instrument. A detailed description of the UVIT and its calibration can be found in \citet{Tandon2017, Tandon2020}. The CCDLAB software package \citep{Postma2017} was used to create the science-ready images with exposure times of 6310.95~s (F148W) and 6268.10~s (F169M). The astrometric calibration was performed using GALEX NUV image \citep{Bianchi2017} and the \gaia data \citep{Gaia2020} as references and the final accuracy was $\sim$ 0.5$\arcsec$.

\begin{figure}[!htb]
\includegraphics[height=0.8\columnwidth]{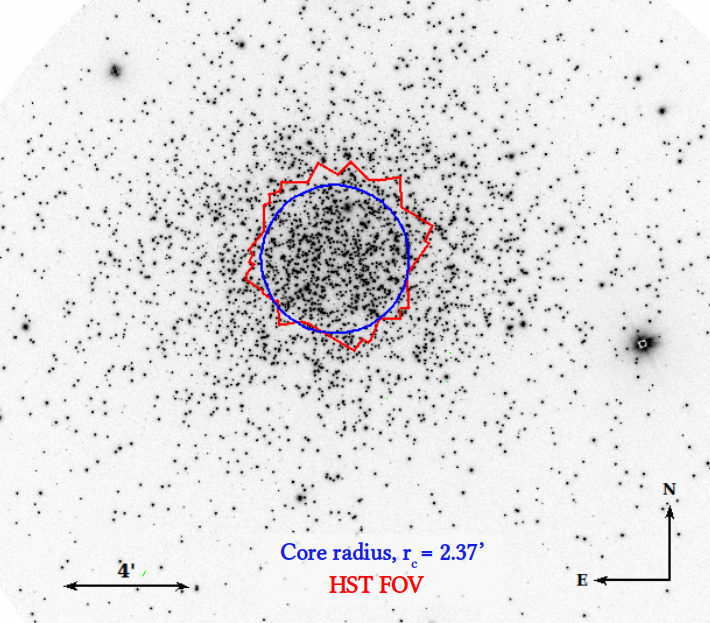}
\vspace{2cm}
\includegraphics[width=0.92\columnwidth]{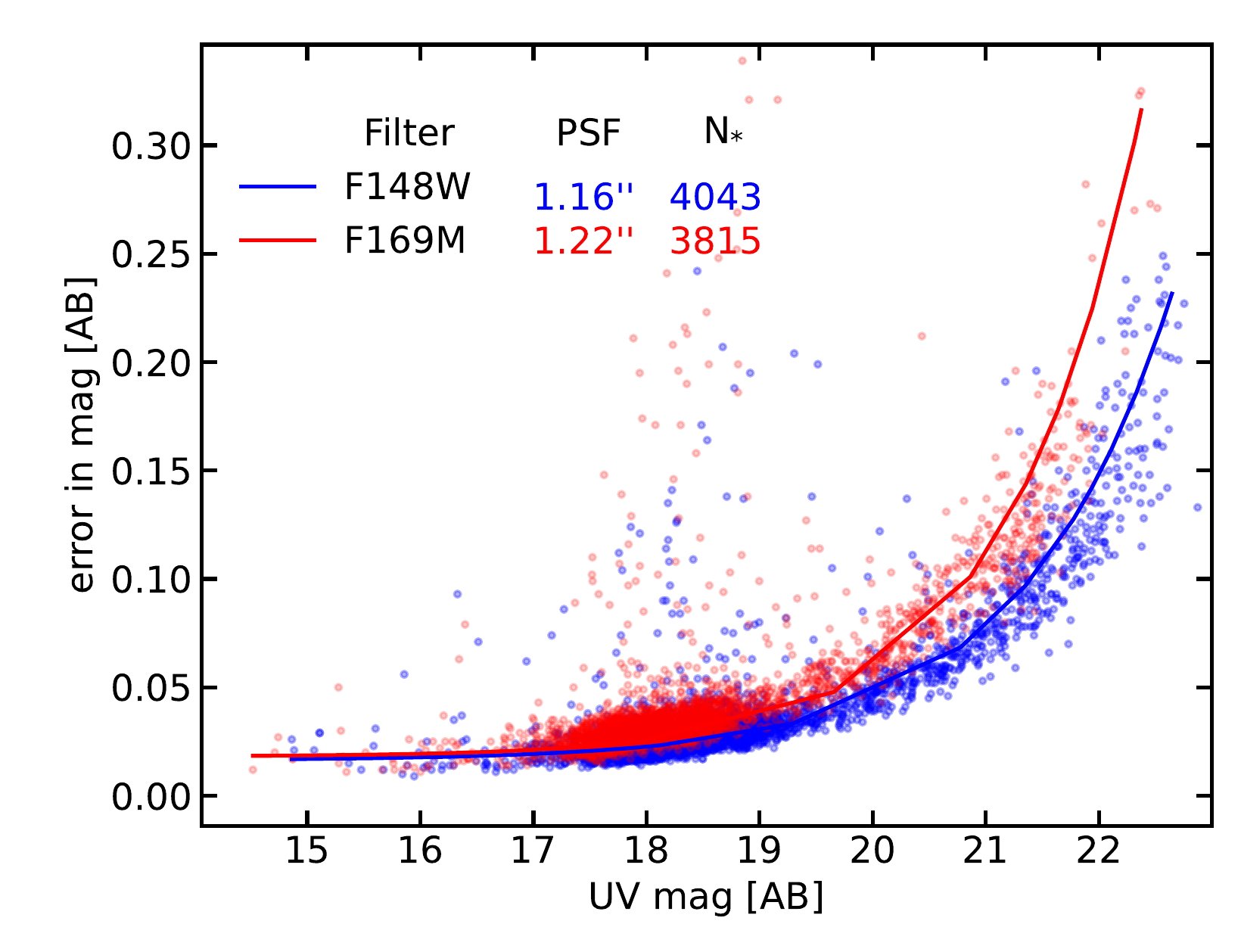}
\caption{Top :UVIT/F148W image of \omcen. Bottom : Plot of PSF-fit errors vs. magnitudes (not corrected for extinction) in the two filters. The filter names, FWHM of the PSF and number of detections with fit error $<$ 0.25\,mag (N$^{*}$) are indicated. The line indicates the median error in each filter.}
\label{omcen_obs_phot}
\end{figure}

We performed PSF photometry on these images as described in \citet{Sahu2022}. 
The source catalog was refined by removing three visibly saturated stars, and those lying at the edge of the UVIT FOV. The final catalog contains only the stars with PSF-fit errors less than 0.25~mag and those detected in both the filters (N = 3697; obtained by matching the co-ordinates within a maximum match radius of 1\arcsec). The UVIT image of the cluster in the F148W filter and the magnitude vs. photometric error plots are shown in Fig.~\ref{omcen_obs_phot}. 

The magnitudes were corrected for extinction by adopting $E(B-V)_{avg}$ = 0.12~mag \harrisp, $R_{V}$ = 3.1, and the Fitzpatrick reddening law \citep{Fitzpatrick1999}. The extinction coefficient values are 0.98~mag and 0.93~mag respectively for the F148W and F169M filters\footnote{Calculated using the York Extinction Solver \citep{McCall2004}}.


\section{Color-magnitude diagrams}

The UV-optical color-magnitude diagram (CMD) of the cluster was constructed by identifying the optical counterparts of the 3689 (out of a total of 3697) FUV-detected sources using the HST-based catalog of \citet{Bellini2017_1} (for $r < $ core radius $r_{c}$ = $\ang[angle-symbol-over-decimal]{;2.37;}$, hereafter inner region) and the catalogs of \citet{Stetson2019} and \citet{Vasiliev2021} (for $\ang[angle-symbol-over-decimal]{;2.37;} < r < \ang[angle-symbol-over-decimal]{;13.5;}$; outer region) as described in Appendix~\ref{cross-match}. The optical magnitudes were converted from Vega to AB system using appropriate conversion factors\footnote{\url{http://waps.cfa.harvard.edu/MIST/BC_tables/zeropoints.txt}}.

\begin{figure*}[!htbp]
\makebox[0.99\linewidth]
{
\includegraphics[scale= 1.4]{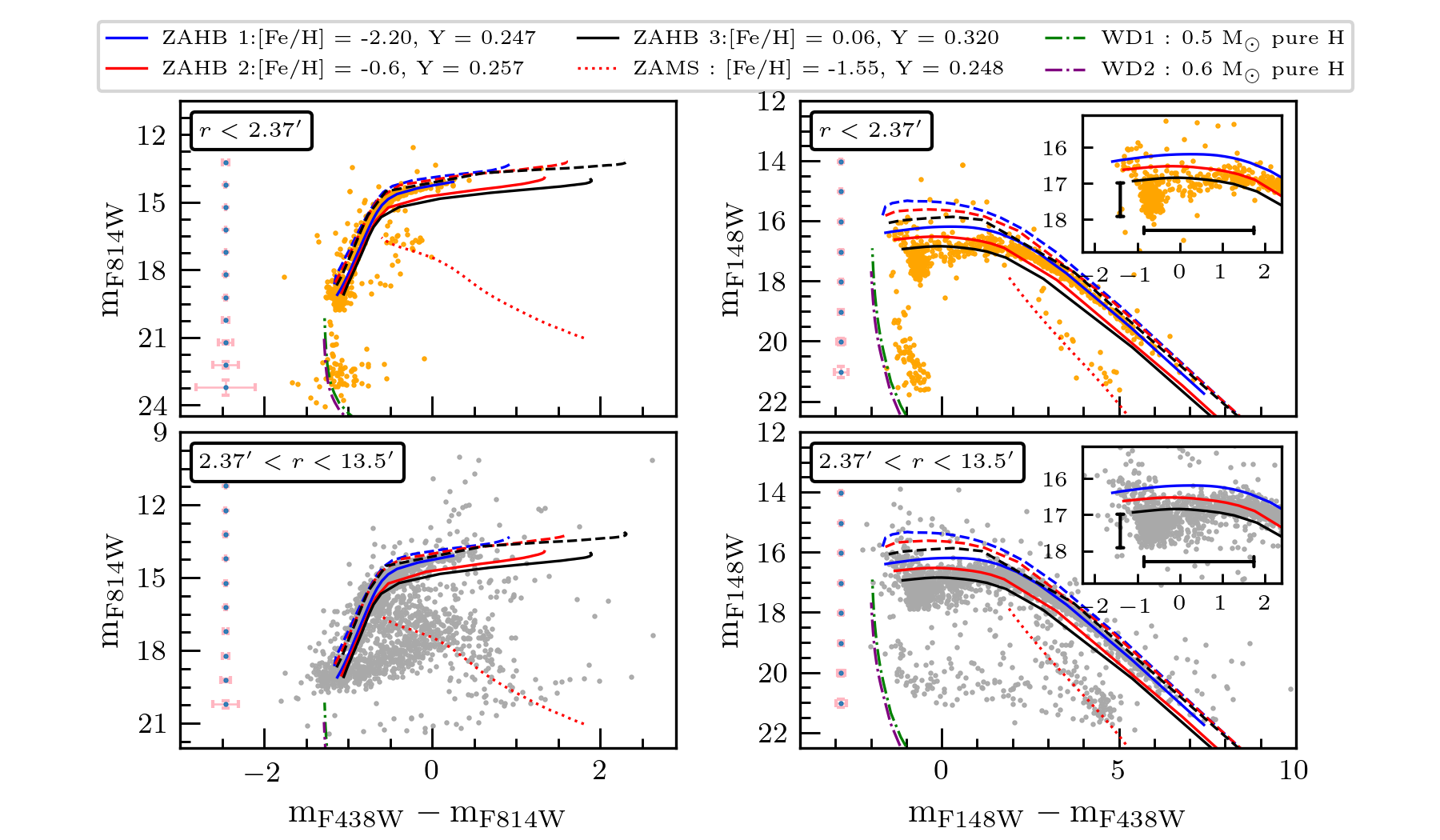}
}
\caption{Theoretical stellar evolutionary models overplotted on the optical and FUV-optical CMDs. The BaSTI ZAHB and TAHB models for different metallicities and He abundances are shown with continuous and dashed lines respectively. The dotted lines represent the BaSTI ZAMS isochrone. The dash dotted lines represent WD cooling sequences. The parameters corresponding to all the models are indicated on the top panel of the figure.}
\label{FUV_cm_sources_models}
\end{figure*}

Fig.~\ref{FUV_cm_sources_models} shows the optical and FUV-optical CMDs of the cluster along with various models and isochrone. For the HB, we used the Bag of Stellar Tracks and Isochrones (BaSTI)\footnote{\url{http://basti-iac.oa-abruzzo.inaf.it/}} \citep{Pietrinferni2021ApJ} theoretical zero-age HB (ZAHB) and terminal-age HB (TAHB) models with \afe=$+$0.4~dex, and mass-loss parameter $\eta=0.3$, where, overshooting is not applied and atomic diffusion effects are included. Three models with the following metallicity and He abundance values were chosen: \feh = $-$2.2~dex, Y = 0.247~dex (metal-poor, He-normal); \feh = $-$0.6~dex, Y = 0.257~dex (metal-intermediate, He-normal) and \feh = 0.06~dex, Y = 0.320~dex (metal-rich, He-enhanced). These choices were based on the values reported in literature and as per the availability in the database. For the BSS sequence, we used the BaSTI zero-age MS (ZAMS) isochrone of age = 0.5~Gyr, with initial mass range $\sim$ 0.5 - 1.5~$M_{\odot}$ corresponding \feh$_{avg}$ = $-$1.55~dex and primordial Y value.
For the WD population, we used two DA spectral type models with pure hydrogen (H) grid and thick H layers with masses 0.5~$M_{\odot}$ and 0.6~$M_{\odot}$ (private comm. with Pierre Bergeron).


The locations of the HBs, BSSs and WDs in the optical CMDs match well with the model predictions (left panels of Fig.~\ref{FUV_cm_sources_models}).  
In the FUV-optical CMDs (right panels), the red HB and BSS sequences lie at the locations expected from the models. It is well known that the hottest HBs, known as blue hook stars (BHKs), appear fainter than canonical models in CMDs \citep{Whitney1998, Dcruz2000}. However, we find that all hot HBs (hHBs) with $m_{F148W} - m_{F438W} \lesssim 2.0$~mag are fainter in the F148W band by about $\sim$~0.5~mag whereas no anomaly is observed in optical CMDs. The WDs too are redder by comparable magnitude only in the FUV-optical CMDs (implying fainter FUV magnitudes). A similar behavior is observed in the UVIT F169M filter (not shown here). Any effect due to the instrument calibration or analysis procedure was ruled out as described in Appendix~\ref{instrument_checks}.

\section{HB simulations}

\begin{figure*}[!htbp]
\centering
{
     \subfloat{\includegraphics[scale= 0.8]{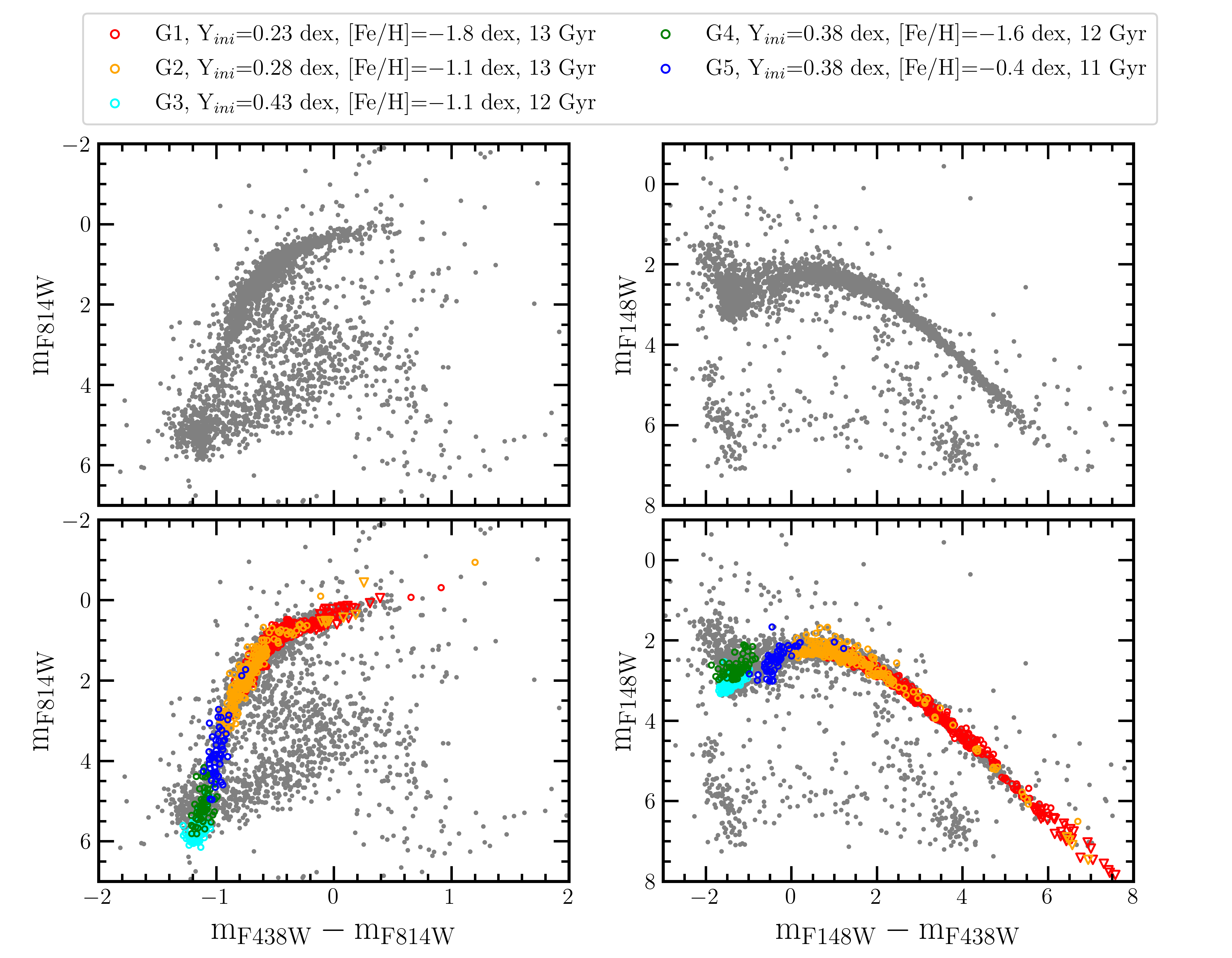}}

    \subfloat{\includegraphics[scale = 0.72]{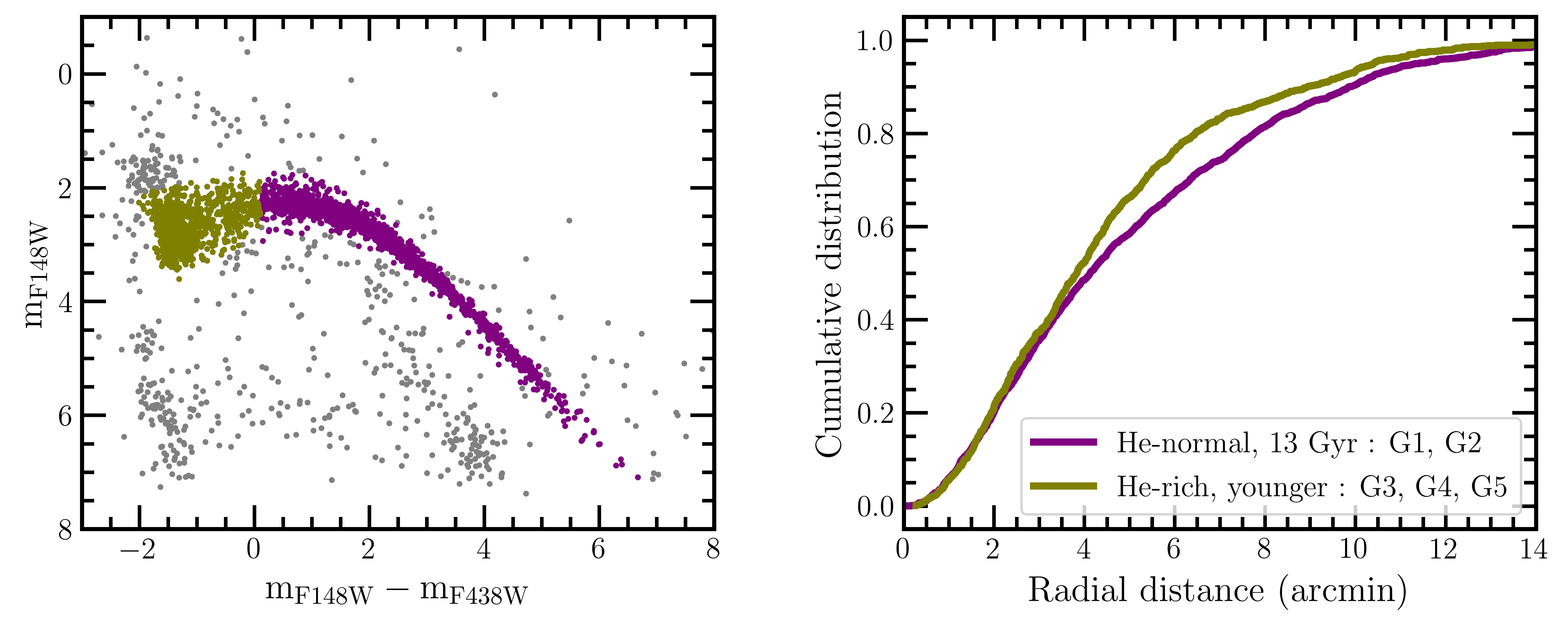}}
}
\caption{HB simulations and radial distributions of subpopulations : The top panels show the observed optical and FUV-optical CMDs and the middle panels show the simulated HB CMDs overplotted on the observed. Parameters suggested from our best fit simulation for subpopulations from G1 to G5 are indicated in the legend above. The triangles in the middle panels denote the simulated RR Lyrae stars. 
The distance modulus and reddening adopted to reproduce the observed CMDs are $(m-M)_{F148W}=15.5$~mag, $E(F148W-F438W)=1.2$~mag, and $(m-M)_{F814W}=14.1$~mag, $E(F438W-F814W)=0.3$~mag, respectively. The bottom right panel shows the radial distribution of observed HB subpopulations where G1 and G2 are considered as a single group and G3, G4 and G5 as another, indicated with purple and olive symbols respectively in the bottom left panel.}
\label{hb_sim_rad_dist}
\end{figure*}

To check if the observed HB distribution originated from the extreme He-enhancement, we produced synthetic CMDs shown in the top and middle panels of Figure~\ref{hb_sim_rad_dist}.
Generally, the CMD synthesis of GCs should simultaneously reproduce both the HB morphology and the main sequence (MS) to red-giant branch stars (RGBs). We can derive reliable stellar parameters and subpopulations' ratios based on this.
However, due to the observational limitations on MS to RGBs in the FUV regime, we performed CMD synthesis only for HBs by referring to the stellar parameters of \citet{JooLee2013} for \omcen who reproduced both sequences simultaneously.
We adjusted three stellar parameters ${\rm Y_{ini}}$, age, and \feh to find the best match to the observations.
The mass-loss parameter was adopted as $\eta=0.5$ for all the stellar populations. 
The detailed descriptions for other parameters and the simulation are summarized in \citet{Chung2017}.
Note that we did not include the evolved phase of HBs (i.e., AGB manqu\'e phase) in the model to avoid the highly uncertain stellar evolution tracks after the He-core depletion.

We assumed five subpopulations to reproduce HB morphologies in two observed CMDs. 
From G1 to G5, the stellar parameters for each subpopulation are indicated above the top panel of Figure~\ref{hb_sim_rad_dist}. The normal He G1 (${\rm Y_{ini}=0.23}$) and slightly He-rich G2 (${\rm Y_{ini}=0.28}$) show reasonable agreements with the observed blue HBs.
In our simulation, the extremely hot HBs mainly originated from G3 and G4 populations with ${\rm Y_{ini}=0.43}$ and $0.38$, respectively.
If we do not change ${\rm Y_{ini}}$, other values of age or metallicity (i.e., extremely old or metal-poor populations) cannot reproduce those HBs.
In addition, as Figure~6 of \citet{JooLee2013} shows, \omcen hosts at least one extreme metal-rich MS to RGB sequence.
To explain this population, we added G5 population with ${\rm [Fe/H]=-0.4~dex}$ and ${\rm Y_{ini}=0.38~dex}$, and this subpopulation matches HBs around $(F148W - F438W) \simeq 0.0$ as well. The fractions of the simulated subpopulations from G1 to G5, adopted based on \citet{JooLee2013}, are 0.49, 0.27, 0.10, 0.07, and 0.07 respectively.

The FUV-optical CMD and the radial distribution of the HB stars belonging to different subpopulations is shown in the bottom panels of Figure~\ref{hb_sim_rad_dist}. Here, the older, He-normal subpopulations G1 and G2 were grouped together (purple symbols in the bottom left panel) and the younger, He-rich G3, G4 and G5 were combined as another sample (olive symbols). The He-rich, second-generation HB stars clearly appear more segregated. The Kolmogorov-Smirnov test returned a $p$-value of $\sim$ 1~$\times$~10$^{-5}$, indicating that the two subpopulations are not drawn from the same distribution.

\section{Characterization of hot populations} \label{sed_fitting}

We used the spectral energy distribution (SED)-fitting technique to characterize the hot stars that showed departure from the BaSTI tracks. The SEDs were constructed and fitted with appropriate models using VOSA (VO SED Analyser; \citealt{Bayo2008}) which generates synthetic photometric points for the chosen filters. The best-fit parameters were estimated by comparing the observed data with synthetic photometry using a $\chi^{2}$ minimization method with the $\chi_{red}^{2}$ given by the relation,

\begin{center}
\begin{equation}
\chi_{red}^{2} = \displaystyle \frac{1}{N-N_{f}} \sum_{i=1}^{N} \Bigg\{ \frac{(F_{o,i}-M_{d}F_{m,i})^{2}}{\sigma_{o,i}^{2}}\Bigg\}
\end{equation}
\end{center}

Here, {\it N} and ${\it N_{f}}$ are the numbers of observed data points and the model parameters fitted respectively, ${\it F_{o,i}}$ is the observed flux, ${\it F_{m,i}}$ is the model flux, ${\it M_{d}} = {\it (\frac{R}{D})^{2}}$ is the multiplicative dilution factor (where \rad and {\it D} are the radius and distance to the star, respectively) and ${\it \sigma_{o,i}}$ is the observed flux error. VOSA calculates two additional parameters, $Vgf$ and $Vgf_{b}$, as {\it visual goodness of fit} indicators, useful when the observational flux errors are underestimated\footnote{See Section 5.1.4 of VOSA documentation for details \\ \url{http://svo2.cab.inta-csic.es/theory/vosa/helpw4.php?otype=star\&what=intro\#}}. We used {\it $D_{avg}$} = 5.426\,kpc \citep{BV2021}.  To account for the extinction, VOSA uses the Fitzpatrick reddening relation. The errors in the fitted parameters were estimated using the statistical approach described in the VOSA documentation.


\subsection {Hot HB stars}
The SEDs of hHBs were fitted using six appropriate models whose parameters, and available ranges are tabulated in Table~\ref{ehb_wd_sed_models}. We employed three approaches aiming to test different aspects. In the first approach, we fitted the SEDs adopting the values listed in the last column of Table~\ref{ehb_wd_sed_models} and by fixing the value of $A_{V}$ to 0.372~mag for $E(B-V)_{avg}$ = 0.12~mag. In the next approach, to check for the effects due to radiative levitation observed in HBs hotter than the Grundahl-jump \citetext{\teff $\sim$~11\,500\,K; \citealt{Grundahl1999}}, we allowed the metallicity parameter to vary up to the solar value keeping $A_{V}$ and the other parameter ranges unchanged. The final approach was meant to check for the effect of differential reddening reported in the cluster \citep{Calamida2005, Bellini2017_2} wherein we included $A_{V}$ as a fit parameter with a range 0.279 - 0.775~mag corresponding to $E(B-V)$ =  0.09 - 0.25~mag, keeping all other parameter ranges as in the last column of Table~\ref{ehb_wd_sed_models}.


\begin{table*}[!hbt]
\caption{Models and parameter ranges adopted to fit the SEDs of hot HB stars and WDs}
\label{ehb_wd_sed_models}
\makebox[0.9\linewidth]
{
\begin{threeparttable}
\begin{tabular}{@{}llll@{}}
\toprule
Model                                & Parameter        & Available range     & Adopted range        \\ 
\bottomrule
\\
\multicolumn{4}{c}{Hot HB stars} \\
\hdashrule[0.5ex]{3.5cm}{1pt}{2pt} & \hdashrule[0.5ex]{3.5cm}{1pt}{2pt} & \hdashrule[0.5ex]{3.5cm}{1pt}{2pt} & \hdashrule[0.5ex]{3.5cm}{1pt}{2pt}  \\
{\multirow{3}{*}{Kurucz ODFNEW/NOVER\tnote{a}}} & \teff & 3\,500 to 50\,000\,K     & 10\,000 to 50\,000\,K  \\
                             & \logg            & 0 to 5\,dex          & 3 to 5\,dex           \\
                                     & \feh       & $-$4.0 to 0.5\,dex     & $-$2.5 to $-$0.5\,dex     \\ 
                                     \cline{2-4}
\multirow{2}{*}{TMAP Grid 2\tnote{b}}        & \teff             & 20\,000 to 150\,000\,K & 20\,000 to 100\,000\,K \\
                                     & \logg            & 4 to 9\,dex          & 4 to 5.5\,dex         \\ 
                                    \cline{2-4}
\multirow{3}{*}{TMAP Grid 4\tnote{b}}         & \teff             & 20\,000 to 150\,000\,K & 20\,000 to 100\,000\,K \\
                                     & \logg            & 4 to 9\,dex          & 4 to 5.5\,dex         \\
                                     & H mass fraction  & 0 to 1              & 0 to 1               \\ 
                                     \cline{2-4}
\multirow{4}{*}{TMAP T\"ubingen\tnote{b}}    & \teff & 30\,000 to 1\,000\,000\,K & 30\,000 to 100\,000\,K \\
                                     & \logg            & 3.8 to 9\,dex        & 3.8 to 5.5\,dex       \\
                                     & H mass fraction  & 0 to 1              & full range           \\
                                     & He mass fraction & 0 to 1              & full range           \\ 
                                     \cline{2-4}
\multirow{4}{*}{\citet{Pacheco2021}} & \teff             & 10\,000 to 65\,000\,K  & full range           \\
                                     & \logg            & 4.5 to 6.5\,dex      & 4.5 to 5.5\,dex       \\
                                     & \feh       & $-$1.5 to 0.0\,dex     & $-$1.5\,dex             \\
                                     & log{[}He/H{]}    & $-$4.98 to 3.62       & full range           \\ 
                                      \cline{2-4}
\multirow{3}{*}{\citet{Husfield1989}}     & \teff             & 35\,000 to 80\,000\,K  & full range           \\
                                     & \logg            & 4.0 to 7.0\,dex      & 4.0 to 5.5\,dex       \\
                                     & Y$_{He}$         & 0.0 to 0.7\,dex      & full range           \\ 
                                     
\hline \\
\multicolumn{4}{c}{WDs}     \\
\hdashrule[0.5ex]{3.5cm}{1pt}{2pt} & \hdashrule[0.5ex]{3.5cm}{1pt}{2pt} & \hdashrule[0.5ex]{3.5cm}{1pt}{2pt} & \hdashrule[0.5ex]{3.5cm}{1pt}{2pt} \\ 
\multirow{2}{*}{Koester\tnote{c}}             & \teff             & 5\,000 to 80\,000\,K   & full range           \\
                                     & \logg            & 6.5 to 9.5\,dex      & full range           \\ 
                                     \cline{2-4}
\multirow{2}{*}{Levenhagen\tnote{d}}          & \teff             & 17\,000 to 100\,000\,K & full range           \\
                                     & \logg            & 7.0 to 9.5\,dex      & full range           \\ \hline
\end{tabular}
\begin{tablenotes}
\item [a] \afe = 0.4~dex; \cite{CK2003}
\item [b] \cite{Werner1999,Werner2003,Rauch2003}
\item [c] \cite{Koester2010}
\item [d] \cite{Levenhagen2017}
\end{tablenotes}

\end{threeparttable}
}
\end{table*}

\begin{figure*}[!htbp]
\makebox[\textwidth]
{
\includegraphics[width=1.1\textwidth, height = 18cm]{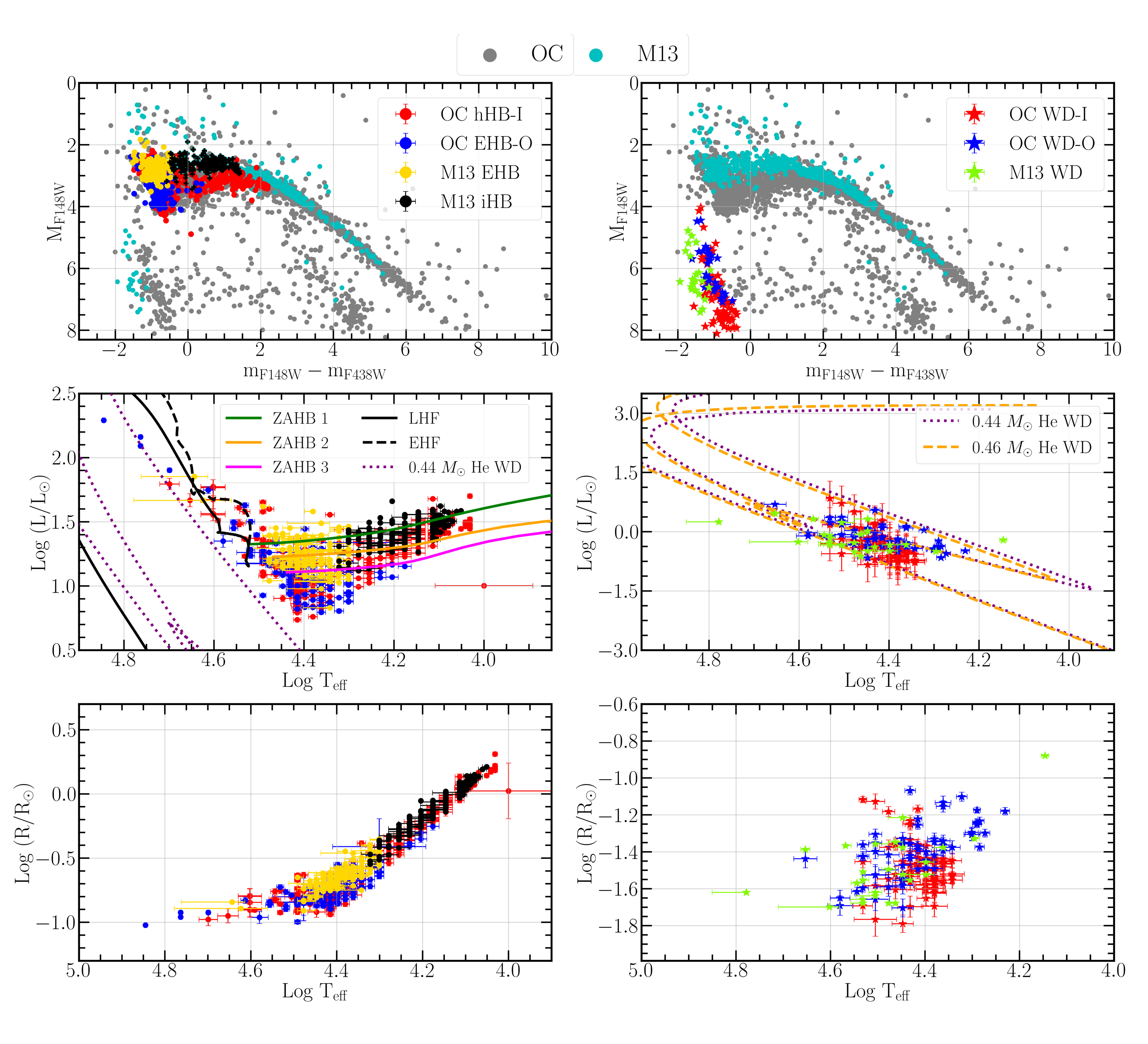}
}
\caption{Comparison of the FUV-optical CMD, H-R diagram, log~\teff vs log~($R/R_\odot$) plots for \omcen (OC) and M13 with focus on the hHB population and WDs. The -I (-O) denotes stars within the inner (outer) region. The parameters corresponding to ZAHB models in the middle left panel are as follows - ZAHB 1 : \feh = $-$2.20\,dex, Y = 0.247\,dex; ZAHB 2 : \feh = $-$0.6\,dex, Y = 0.257\,dex; and ZAHB 3 = \feh = 0.06\,dex, Y = 0.32\,dex. LHF is a late hot flasher model with [M/H]=$-$1.018\,dex, Y = 0.250\,dex, M$_{ZAHB}$ =  0.490\,$M_{\odot}$ and EHF is an early hot flasher model with [M/H] = $-$1.90\,dex,  Y = 0.247\,dex and  M$_{ZAHB}$ = 0.502\,$M_{\odot}$.} 
\label{m13_ocen_ehb_wd_sed_fit_comp}
\end{figure*}

The top left panel of Fig.~\ref{m13_ocen_ehb_wd_sed_fit_comp} shows the FUV-optical CMD highlighting the sample of hHBs analyzed. Among the inner region stars, we chose all the 421 HBs bluer than the knee point in the CMD (at color $\sim$ 2.0~mag), which are shown with red circles and denoted as OC hHB-I. For these stars, the UVIT photometry in two filters was combined with the photometric data in 18 HST WFC3/UVIS filters from \citet{Bellini2017_1}. Among the outer sources, we chose the 150 extreme HB (EHB; \teff $\gtrsim$ 20\,000\,K) stars with confirmed cluster membership obtained through a cross-match with the sample of \citet{Latour2018} (shown with blue circles and denoted as OC EHB-O). The SEDs of these stars were constructed by complementing the UVIT photometry with the data in five optical filters from the catalog of \citet{Stetson2019}. We thus derived the physical parameters for a total of 571 hHBs, using the three approaches mentioned above. Since the results were found not to differ much, the discussions below are based on the first approach (also shown in the figure).

Good fits, with $Vgf_{b}$ $<$ 15 \citep{Rebassa2021}, were achieved for about 97\% of the hHBs. The middle left panel of Fig~\ref{m13_ocen_ehb_wd_sed_fit_comp} shows the Hertzsprung–Russell (H-R) diagram for these stars along with the same ZAHB models as in Fig.~\ref{FUV_cm_sources_models}, early and late hot-flasher (EHF \& LHF) models \citep{Cassisi2003} and a 0.44~$M_{\odot}$ He-core DA-type WD model with  $Y_{ini}$ = 0.4~dex, $Z$ = 0.0005~dex progenitor \citep{Althaus2017}. Most of the stars cooler than log~\teff $\sim$ 4.2 lie within the range of the ZAHB models. A significant fraction within log~\teff $\sim$ 4.2 to 4.5 are fainter than any of the ZAHB models. Their luminosity decreases with \teff reaching a minimum at log~\teff $\sim$ 4.4 which then increases further. The hHBs within the log~\teff $\sim$ 4.5 to 4.65 are lying on the EHF and LHF tracks. The ones hotter than log~\teff $\sim$ 4.6 follow the 0.44~$M_{\odot}$ WD model. In the bottom left panel, the log~(R/R$_\odot$) of hHBs are plotted as a function of the log~\teff, where the low-luminous hHBs of \omcen are found also to be smaller in size.

\subsection{WDs}
The top right panel of Fig.~\ref{m13_ocen_ehb_wd_sed_fit_comp} shows the FUV-optical CMD highlighting the WD sample chosen for SED-fitting. There are 77 (68) WD candidates from the inner (outer) region represented by red (blue) star symbols and denoted as OC WD-I (OC WD-O), whose SEDs were fitted using models and parameters tabulated in Table~\ref{ehb_wd_sed_models}. While inspecting the SEDs, we found about 8 outer WDs showing UV excess. Their SEDs could not be fit with single WD model and hence were not considered further. We also excluded the fits with $Vgf_{b}$ $>$ 15. The SED parameters for the rest of the sample (135 stars) are shown in the H-R diagram on the middle right panel of Fig~\ref{m13_ocen_ehb_wd_sed_fit_comp} along with two He-core models of masses 0.44~$M_{\odot}$ and 0.46~$M_{\odot}$ with  He-rich ($Y_{ini}$ = 0.4~dex), $Z$ = 0.0005~dex progenitors \citep{Althaus2017} \footnote{Note that the adopted He-core WD models do not result from binary evolution \citetext{see \citealt{Althaus2017} for details}.}. Most stars lie in between the range of the models. The bottom right panel shows the (log~$R/R_\odot$) of the WDs as a function of the log~\teff, where the two populations of the WDs match well, except for a small number of cooler and larger WDs in the outer \omcen.

\section{Comparison of hot populations in Omega Centauri and M13}

To further understand the peculiarities seen in the properties of hot stellar populations of \omcen, we compared them with those of another massive GC, M13 (or NGC~6205). M13 has \feh = $-1.53$~dex \harrisp similar to the \feh$_{avg}$ of \omcen with $\Delta$Y$_{max}$ $\sim$~0.05~dex \citep{Dalessandro2013_m13, Milone2018} and age $\approx$ 13~Gyr \citep{Denissenkov2017}. M13 was also observed using UVIT as part of GlobULeS and the detailed study has been carried out by Kumar et al. 2022 \citetext{in preparation}. We obtained the final photometric data and the SED-fit results of HBs (in the range 11\,500\,K $\leq$ \teff $\leq$ 20\,000\,K; iHBs) and WDs from the authors. These are highlighted in yellow, black and light green symbols among the rest of the M13 FUV sources (cyan) on the FUV-optical CMDs (top panels of Fig.~\ref{m13_ocen_ehb_wd_sed_fit_comp}). The \omcen populations with $m_{F148W} - m_{F438W} \lesssim 2$ are clearly fainter in the FUV, when compared to those of M13, whereas the locations of redder stars match well. 

The \omcen hHBs are on an average less luminous than their counterparts in M13 and the latter also fall within the range of the ZAHB models as seen in the left middle panel. 
In the log~\teff vs log~($R/R_\odot$) plot, a few of the \omcen hHBs (mostly near log~\teff $\sim$ 4.4) are found to have smaller radii than their M13 counterparts. This could explain their lower luminosities. The above may also imply that a fraction of hHBs in \omcen are less massive than those of M13. 
The WDs in both clusters occupy similar positions in the H-R diagram and the log~\teff vs. \rad plot on the middle and bottom right panels respectively.  


\section{Discussion and Summary} \label{summary}

The first comprehensive FUV study of \omcen reveals that the HBs bluer than the knee point in the FUV-optical CMD and the WDs are fainter in the FUV by about $\sim$ 0.5\,mag than canonical expectations and in comparison with populations of another cluster having similar properties, namely, M13. \citet{Monibidin2012}, by deriving color-temperature relations, found analogous results uniquely for \omcen hHBs in the $U$ band while detecting no anomaly in the $B$ and $V$ bands. The authors were unable to fully account for this. 

We simulated the HB and compared it with observations to estimate He-enhancements, metallicities, and ages of subpopulations. We find that at least five subpopulations with three He-rich ones are needed to explain the observed HB CMDs. As well known, it would be challenging to determine metallicity and age using HB CMDs only. However, in terms of HB morphology, we conclude that a considerable amount of He-enhancement is inevitable to explain the hHBs in FUV and optical CMDs. The fainter FUV magnitudes of hHBs could be due to the sensitivity of FUV bands to the $Y_{ini}$ range \citep{Chung2017}. The derived parameters of the subpopulations are also comparable to the studies by \citet{JooLee2013}. We find the He-rich younger subpopulations ($\sim$ 24\%) to be radially more segregated than the He-normal older subpopulations ($\sim$ 76\%), which is expected if the second-generation stars form from the ejecta of intermediate-mass asymptotic giant branch stars \citep{D'ercole2008}. \citet{Bellini2009} reported similar results for He-rich MS stars. The ranges in age, metallicity and He content that are needed to fit the observed HB distribution provide constraints on the \omcen formation models.

The properties of hHBs in \omcen and M13 are in general comparable, except for a small fraction of low luminous ones in \omcen. Through spectroscopic measurements, \citet{Monibidin2011} and \citet{Latour2018} reported a mean mass lower than canonical expectations for EHB stars in \omcen  (0.38\,M$_{\odot}$) and could not explain this conundrum.

The WDs in \omcen and M13 have similar physical parameters. However, unlike \omcen, M13 is not known to host extreme He-rich stars that can form He-core WDs from single stellar evolution. \citet{Chen2021} suggested that the bright WDs in M13 are the result of slow cooling due to the residual hydrogen-burning on the C/O WD surface. Hence, it is possible that some of the FUV-detected WDs in \omcen are such slowly cooling C/O WDs. Photometric observations of WD pulsations in the future can shed more light in this direction \citep{Althaus2017}.

\acknowledgments

We thank the anonymous referee for the helpful comments and suggestions. We thank Andrea Bellini for making the HST photometric data available. We are grateful to Pierre Bergeron \& Leandro Althaus for making available the WD models. DSP thanks Ranjan Kumar for sharing the results of M13 and Simone Zaggia for the useful discussions. This publication utilizes the data from {\it AstroSat} mission's UVIT, which is archived at the Indian Space Science Data Centre (ISSDC). The UVIT project is a result of collaboration between IIA, Bengaluru, IUCAA, Pune, TIFR, Mumbai, several centers of ISRO, and CSA. This research made use of VOSA, developed under the Spanish Virtual Observatory project supported by the Spanish MINECO through grant AyA2017-84089. AS acknowledges support from SERB Power fellowship.
C.C. acknowledges support from the National Research Foundation of Korea (2022R1A2C3002992, 2022R1A6A1A03053472).
NWCL gratefully acknowledges the generous support of a Fondecyt Iniciaci\'on grant 11180005, as well as support from Millenium Nucleus NCN19-058 (TITANs) and funding via the BASAL Centro de Excelencia en Astrofisica y Tecnologias Afines (CATA) grant PFB-06/2007.  NWCL also thanks support from ANID BASAL project ACE210002 and ANID BASAL projects ACE210002 and FB210003. AARV acknowledges the funding from ANID, Millennium Science Initiative, ICN12\_009.


\appendix

\section{Cross-match of UVIT detections with different catalogs} \label{cross-match}

\subsection{UVIT-\hst cross-match}

The optical counterparts of the FUV detections were identified by cross-matching those within the core radius of the cluster ($r_{c}$ = $\ang[angle-symbol-over-decimal]{;2.37;}$) with the HST dataset from \citet{Bellini2017_1}. This astrophotometric dataset was available in 18 WFC3/UVIS bands and 8 WFC3/IR bands for 478,477 stars. However, we used only the WFC3/UVIS data since the IR photometry did not include the information of the saturated stars, many of which belonged to the HB. The saturated stars also did not have proper motion (PM) measurements in the catalog. Hence, we did not perform a filtering of the sources based on cluster membership. A direct cross-match of the UVIT and HST catalogs would result in many spurious identifications because of the differences in the spatial resolution and astrometric accuracy. Hence, we chose a subset of stars from the HST dataset which included only the UV-bright stellar populations such as HB, post-HB, BSSs and WDs. The HST photometry in \citet{Bellini2017_1} was measured with three methods, each of which worked best in different magnitude regimes. Following the same selection criteria as the authors, we used the results of method one photometry for HB, pHB and BSS stars and those of method two for WDs. The UV color-magnitude plane $m_{F275W} - m_{F336W}$ vs. $m_{F275W}$ was used to select the HB, pHB and BSS stars and the WDs were selected from the $m_{F438W} - m_{F606W}$ vs. $m_{F438W}$ plane. This HST subset was cross-matched with the UVIT catalog with a maximum match radius of $\ang[angle-symbol-over-decimal]{;;0.7}$, using TOPCAT and about 963 stars were found to have unique counterparts. We also manually verified that all the cross-identifications were accurate. A counterpart could not be identified correctly for one star at R.A. = 201.64700$^{\circ}$ and Decl. = $-$47.50401$^{\circ}$ in the UVIT catalog (it is not included in HST catalog although it is visible in F555W image). So this star is excluded.


\subsection{UVIT-ground data-Gaia EDR3 cross-match}

There were several UV-bright stars lying outside the HST FOV. In order to analyze these stars and plot them in the UV-optical CMD, we used the ground-based optical dataset in the $UBVRI$ filters from \citet{Stetson2019} and the \gaia EDR3-based catalog from \citet{Vasiliev2021}. The latter also included cluster membership probability estimates based on PM measurements. For the cross-match, we first created a subset of stars from the ground-based catalog, selecting only the population expected to be bright in UV. This subset was matched with the UVIT detections with a maximum cross-match radius of $\ang[angle-symbol-over-decimal]{;;0.7}$. The number of stars common in both was about 2725. In order to identify the cluster members among them, we matched this set further with the \gaia-based catalog of \citet{Vasiliev2021}. About 1771 stars were found to have membership probability more than $0.5$. However, there were about 803 UV-bright stars which were not included in the \gaia EDR3 catalog. Hence, their membership status is unknown. Finally, to plot all the sources in similar color-magnitude plane,  we transformed the Johnson-Cousins $B$, $V$, and $I$ magnitudes of the stars in the outer region to the corresponding HST WFC3/UVIS filters (namely, $F438W$, $F606W$, and $F814W$), using the equations from \citet{WEHarris2018}. 


\section{Checks for effects due to instrument calibration or analysis procedure} \label{instrument_checks}

\begin{figure*}[!htbp]
\centering
     \subfloat{\includegraphics[scale=0.9]{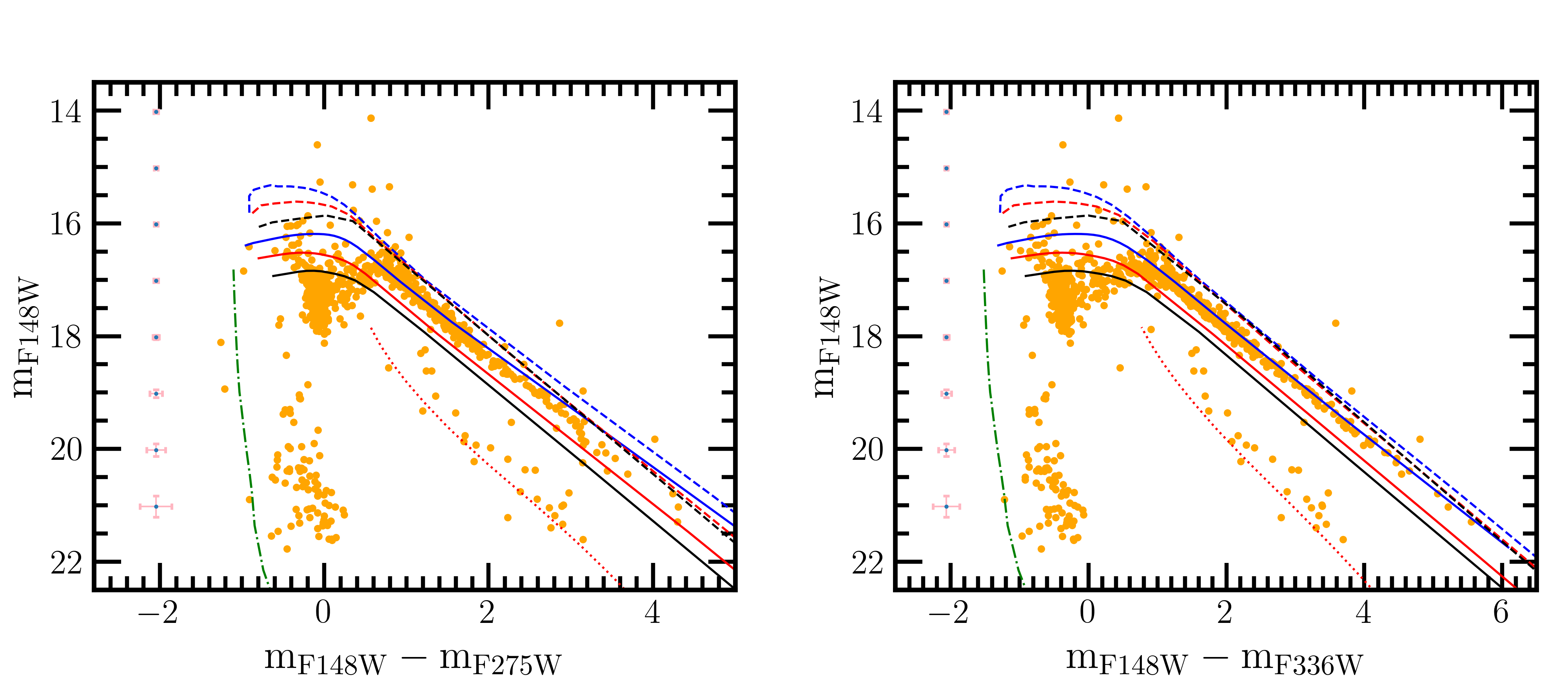}}\\
     \subfloat{\includegraphics[scale= 0.75]{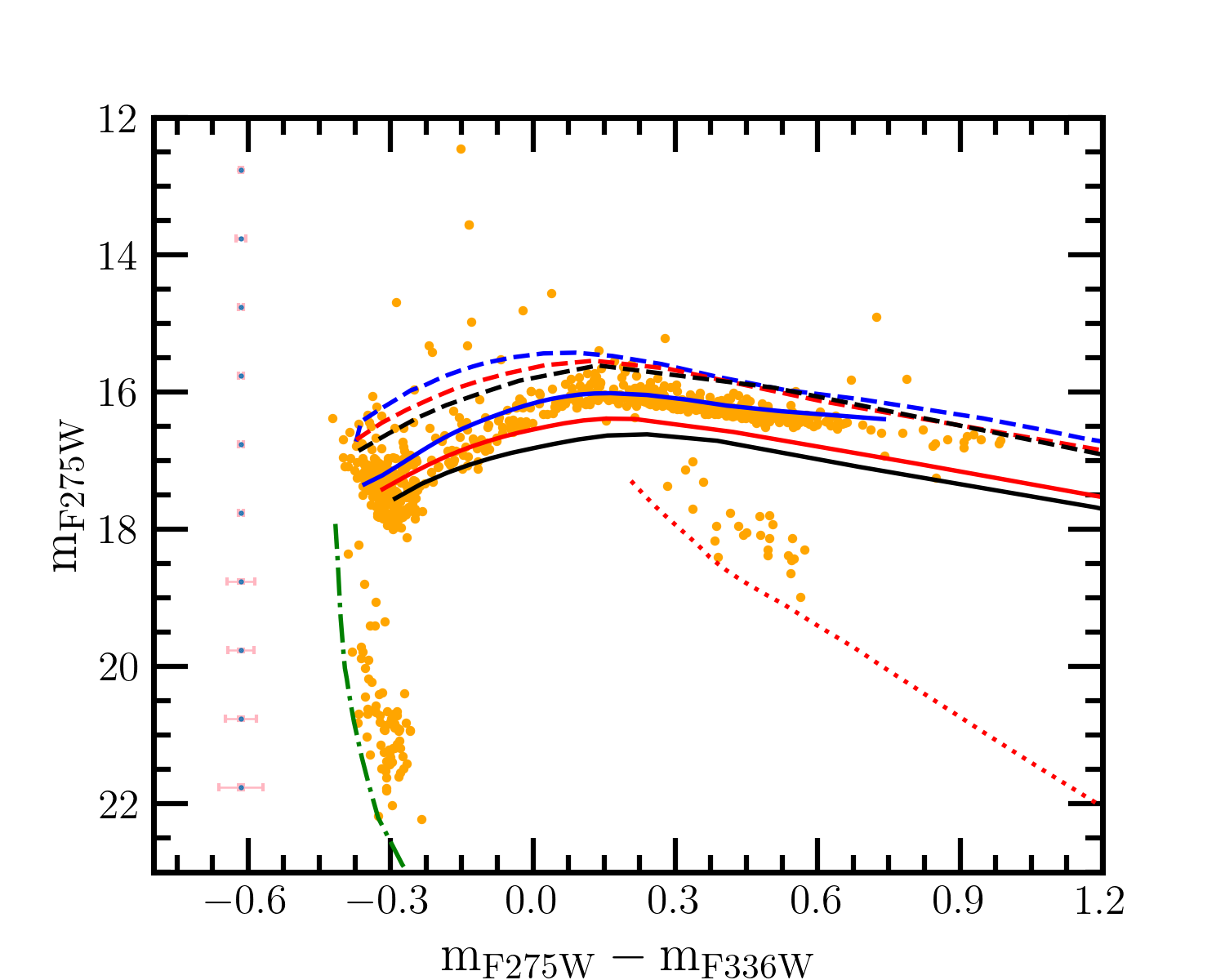}}
     \subfloat{\includegraphics[scale = 0.75]{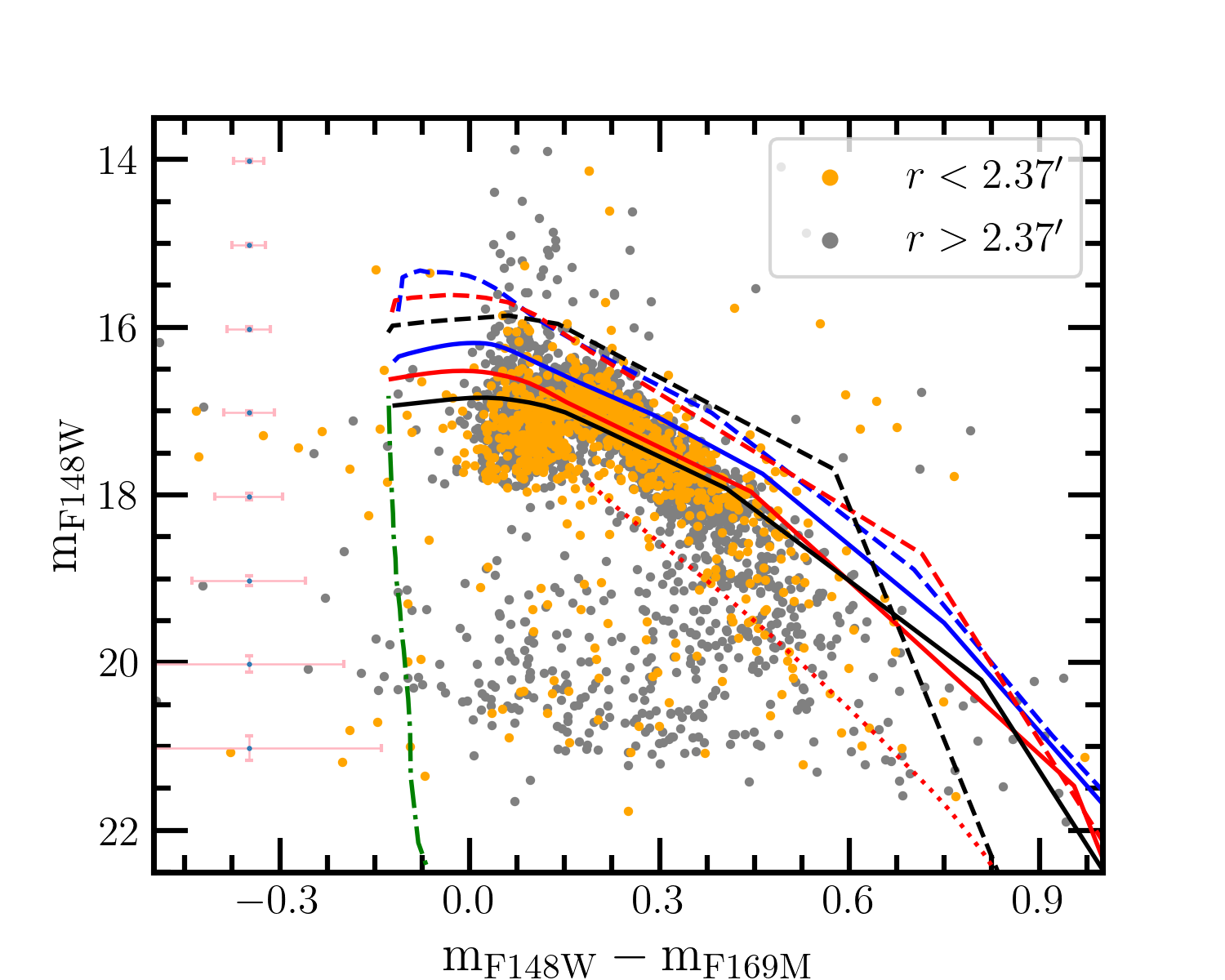}}

\caption{Top panels : The UVIT/F148W$-$HST/F275W vs HST/F275W and UVIT/F148W$-$HST/F336W vs HST/F336W CMDs showing the UVIT-HST common detections (963 stars), along with the theoretical stellar evolutionary models shown in Fig.~\ref{FUV_cm_sources_models}. Bottom left : The F275W$-$F336W vs F275W CMD with the same 963 stars as above, is shown. Bottom right : The CMD consisting of all the sources detected in both the UVIT F148W and F169M filters, along with evolutionary models. }
\label{fuv_nuv_cmds}
\end{figure*}

We checked for various aspects that could possibly result in the bias observed in the FUV-optical CMDs. Firstly, effects due to UVIT instrument-related aspects such as changes in calibration, sensitivity and the slope of the transmission window were inspected. These were ruled out as magnitudes obtained from the recent observations of FUV-bright sources in the secondary calibration source, open cluster NGC~188, and a previously studied GC NGC~2808, were consistent with earlier estimates. Next, we examined effects due to data reduction and analysis procedures. We obtained similar magnitudes (within $\sim$ 0.1~mag) with science-ready images produced using CCDLAB and the official UVIT L2 pipeline, ruling out any issue due to the data-reduction pipeline. Photometric analysis with IRAF was checked independently and found to be consistent. Lastly, we looked for possible changes introduced due to the transformation of optical magnitudes from the Vega to AB system. The bias was found even when the filter with the smallest transformation factor (F606W) was used, ruling out this possibility. In the top panels of Figure~\ref{fuv_nuv_cmds}, we show the CMDs constructed using the UVIT F148W and the HST F275W (NUV) and F336W (UV wide) filters, consisting of the 963 UVIT-HST common detections. We find that the hHBs and WDs show a deviation from model predictions in these CMDs also. Additionally, the CMD constructed using only the HST filters F275W and F336W, with the data from \citet{Bellini2017_1}, also shows the deviation, supporting our analysis (bottom left panel of Fig.~\ref{fuv_nuv_cmds}). In the bottom right panel of Fig~\ref{fuv_nuv_cmds}, we show the CMD with sources detected in both the UVIT FUV filters.

\newpage

\bibliography{ngc5139}{}
\bibliographystyle{aasjournal}

\end{document}